\documentstyle[]{l-aa}

\def\n2188{NGC\,2188}
\def\ha{H$\alpha$}
\def\kms{\rm \,km\,s^{-1}}
\def\stot{\sigma_{\rm tot}}
\def\sgas{\sigma_{\rm gas}}
\def\rigm{\rho_{\rm IGM}}

\begin{document}

\title{What perturbs \n2188?\thanks{Partly based on observations
obtained at ESO/La Silla (Chile)}}
\author{H. Domg\"orgen\inst{1}, M. Dahlem\inst{2}{\thanks{Affiliated
with the Astrophysics Division in the Space Science Department of ESA}}, 
R.-J. Dettmar\inst{3}}
\institute{Sternwarte der Universit\"at Bonn, Auf dem H\"ugel 71,
D--53121 Bonn, Germany \and 
Space Telescope Science Institute, 3700 San Martin Drive, Baltimore, MD 21218,
USA \and
Astronomisches Institut der Ruhr-Universit\"at Bochum, Universit\"atsstr. 150,
D--44780 Bochum, Germany}
\date{Received,Accepted}
\thesaurus{03(09.02.1; 11.08.1; 11.09.1 NGC\,2188; 11.09.3; 11.09.5; 11.09.4)}
\offprints{H. Domg\"orgen}
\maketitle

\begin{abstract}{
VLA HI observations of the dwarf irregular galaxy \n2188 show
that the gas and stars have spatial distributions which
are substantially different.  One end of the optical disk is strongly
gas deficient, while neutral gas extends into the halo over distances 
of more than 2 kpc from the midplane. This and the peculiar velocity 
field suggest that \n2188 is a perturbed system, although
it is not obviously an interacting galaxy.  

In addition, \n2188 is remarkable for its interstellar disk-halo connection.
An \ha\/ image of \n2188 shows the presence of spectacular 
features of ionized gas extending from a large star-forming complex
up to 500 pc into the halo.  Also, peculiar HI 
filaments and at least one superbubble are present in \n2188.
}
\end{abstract}
\section{Introduction}

Dwarf irregular galaxies are ideal laboratories to study the
feedback processes between star formation and the interstellar
medium (ISM).  Due to their low mass and hence low gravity,
young massive stars are able to blow large holes into the surrounding ISM 
(e.g. Holmberg II, Puche et al. 1992)
and galactic winds are thought to play a major role in the
evolution of dwarf galaxies (Marlowe et al. 1995, and references therein).  
The structures produced are 
longlived since differential rotation is hardly important.  

Powerful tools to study the impact of massive stars on
the ISM are H$\alpha$ images (showing the distribution
of star-forming regions) and HI imagery (showing the superbubble
expansion).  The galaxies NGC\,4449 (Bajaja et al. 1994), 
IC\,10 (Shostak \& Skillman 1989), 
and Sextans A (Skillman et al. 1988) are just some 
examples that HI synthesis maps are also necessary to understand the
full nature of dwarf irregular galaxies (Skillman 1994).

\n2188 is an edge-on (i$\sim86^{\circ}$) Magellanic type
irregular galaxy (Tully 1988).  Its absolute blue magnitude, 
$M_B \sim$ --17.87 mag which is based on a distance  
of 7.9 Mpc (Tully 1988)
indicates that its size is somewhere between those of 
dwarf galaxies and normal spirals.  Although \n2188
was not detected in CO (Israel 1995), the measured colour 
index, $U-B$ = $-$0.21, hints at the presence of a population of young stars.  

As part of a project to investigate the interstellar disk-halo connection
of nearby galaxies we obtained observations of \n2188.
Here we present a detailed study of \n2188 based on
a H$\alpha$ image as well as high resolution HI data.
To our surprise we discovered large scale asymmetries
between the optical and radio image of the galaxy.  Additionally, huge 
gas structures extending from the disk into the halo were found.

\section{Observations and data reduction}

\subsection{HI line data}

HI data of 
\n2188 were collected with the VLA\footnote{The VLA is a facility of the 
National Radio Astronomy Observatory, which is operated by Associated 
Universities, Inc., under contract with the National Science Foundation} 
synthesis telescope during two observing runs of 5 h each in October 1994 
(CnB array) and a third run of 3.5 h in January 1995 (DnC array), 
filling in short baselines in order to obtain good sensitivity for 
extended emission. 

The individual data sets were calibrated in a standard fashion with the 
AIPS software package. 3C147 was used as primary flux calibrator and 
0616$-$349 for the phase calibration. After calibration all data of both 
arrays were combined. The final data set has a resolution of $11\farcs9 
\times 11\farcs2$. 
%and the rms noise in each channel is 1.3 mJy/beam.  
After the continuum subtraction the maps showing line emission were
CLEANed. Using uniform weighting, $512\times512$ pixel maps with a 
pixel size of $3\farcs5$, a resolution of $12''$, and an rms noise of 
1.3 mJy/beam were produced.  Applying natural weighting we also created 
maps with a resolution of 25$''$ and an rms noise of 0.3 mJy/beam. 
Both map versions are used in the analysis: 
the uniform weighted maps for investigations of small scale structures, 
and the natural weighted maps for investigating extended structures. 
The velocity resolution of our data is 5.18 $\kms$.

\subsection{Optical narrow band imaging}

H$\alpha$ images of \n2188 were obtained in Feb. 1993 with 
the ESO 2.2m telescope on La Silla, using EFOSC\,2 (Melnick et al., 1989) with
a $1024\times1024$ pixel Thompson CCD chip. The spatial 
scale is $0\farcs34$/pixel; the seeing during our observations was 
$\simeq0\farcs9$, which gives a linear resolution of 34 pc. The 
integration times were $2\times30$ min with the H$\alpha$ filter 
\#694, which has a  $FWHM$ of  $61\ {\rm \AA}$ and a central wavelength 
of $\lambda_{\circ} = 6557\ {\rm \AA}$, and 10 min with Gunn $r$.

The data reduction was performed using the IRAF software package. 
Bias and dark were subtracted and 
gain variations were removed using dome flatfields. 
We performed the continuum subtraction using the scaled R-band 
image (e.g., Dettmar 1990).

\section{Data analysis and results} 

\subsection{HI line emission from \n2188}

Using the channel maps, a ``moment analysis'' was performed, applying 
the so-called ``unsharp masking'' technique with a 3-$\sigma$ significance
level for blanking.  

\subsubsection{HI content and distribution}

The total HI line flux of \n2188 in our data is $S_{HI} = 20.4\pm0.6$ 
Jy $\kms$. Assuming the typical case of small optical depth, this
corresponds to an HI gas mass of $3\times 10^8$ M$_{\odot}$. Using 
the Parkes telescope Reif et al. (1982) found $S_{HI} = 32.9$ Jy 
$\kms$ for \n2188.  Thus, $\sim$38\% of the total flux might be missed 
by the VLA measurements due to missing short spacings, or 
\n2188 could be surrounded 
by a halo of low surface brightness HI gas below our current detection 
limit of $1.6\times10^{20}$ cm$^{-2}$.
However, this 
discrepancy is uncertain, because the Reif et al. measurements are of
low S/N.

\begin{figure}
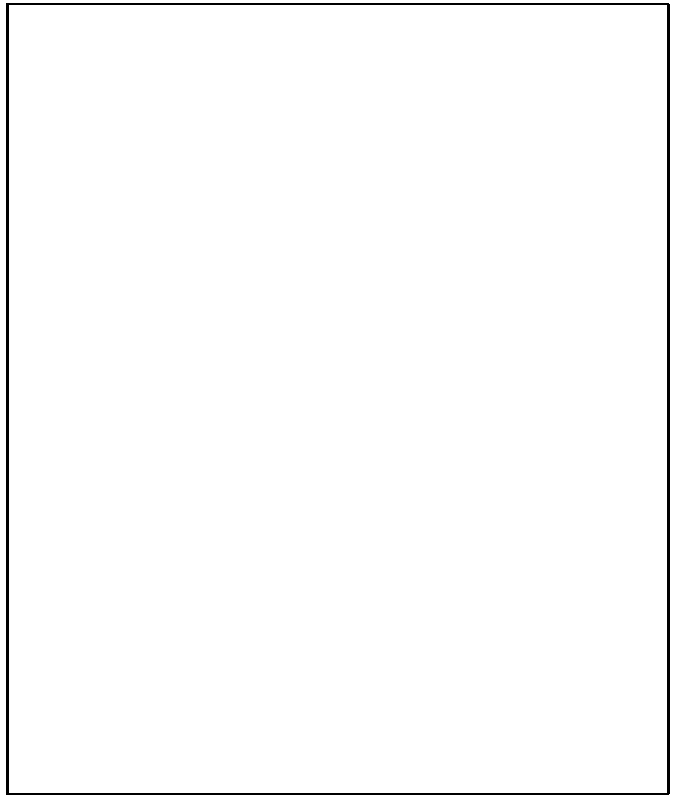

\picplace{10.5cm}
%\def\epsfsize#1#2{1.0\hsize}
%\centerline{\epsffile{hi+ha.ps}}
\caption[]{\ha\/ image of \n2188 with overlying HI contours.
Contour levels are: $2.5, 7.5,..., 47.5 \times 10^{20}$ cm$^{-2}$.}
\end{figure}

Superimposed on the H$\alpha$ frame of \n2188 we present in Figure 1
contours of the total HI line flux density. The disk of \n2188 
is well defined by a high surface brightness ridge.  Column densities 
reach peak values of $4.8\times10^{21}$ cm$^{-2}$ and drop rapidly along 
the minor axis.
A comparison of the HI contours and the \ha\/ distribution shows an 
overall correspondence of the HI high surface brightness disk and the 
star forming disk.
Note that the position angle of the \ha\/ disk does not stay 
constant over the whole disk.  Instead we find the disk to be slightly 
bent to the east near both ends of the H$\alpha$ emission distribution,
leading to a ``banana''-shape.

The disk is surrounded by an extended envelope of neutral gas which  
is distributed asymmetrically with respect to the 
disk:  HI emission corresponding to a column density of 
$2.4\times10^{20} $ cm$^{-2}$ can be found 
$\sim$2.7 kpc east of the midplane and $\sim$1.6 kpc to the west (Fig. 1).

Intensity profiles along the minor axis of \n2188 can
be approximated with a Gaussian component (representing the beam-smeared 
disk emission) plus two wings of emission extending further out. In order 
to get a rough estimate of the percentage of emission coming from this 
extended component we produced a representative intensity profile along 
the minor axis in the declination range $-34^{\circ}\,04'\,33'' > \delta 
> -34^{\circ}\,06'\,53''$ (thus avoiding contributions from the peculiarly 
bent outer ends of the disk). 
Subtracting a Gaussian profile from the resulting emission distribution 
we find that the residual corresponds to $\sim30\%$ of the emission.  

\begin{figure}
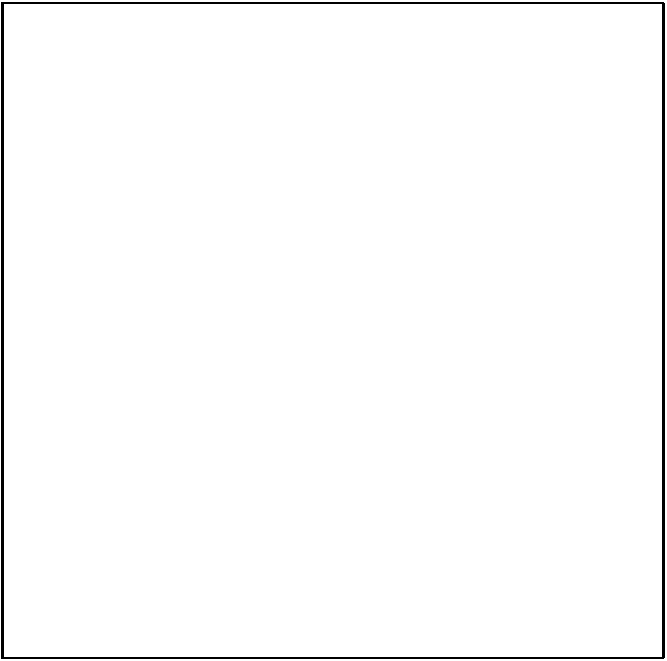

\picplace{8.7cm}
%\def\epsfsize#1#2{1.0\hsize}
%\centerline{\epsffile{gasp+hi.ps}}
\caption{Superposition of the HI data of \n2188 on a digitized
POSS plate.
Contour levels are: $2.5, 7.5,..., 47.5 \times 10^{20}$ cm$^{-2}$.}
\end{figure}

Figure 2 shows a superposition of the HI map of \n2188 on a digitized
POSS plate. Here one can note another peculiarity of this galaxy.  While 
there is plenty of neutral gas extending perpendicular to the disk in 
places where we do not find optical continuum emission, we find optical 
continuum (i.e., stellar light) in the disk plane at the northern end of 
the disk where no HI emission could be detected.  

\begin{figure}
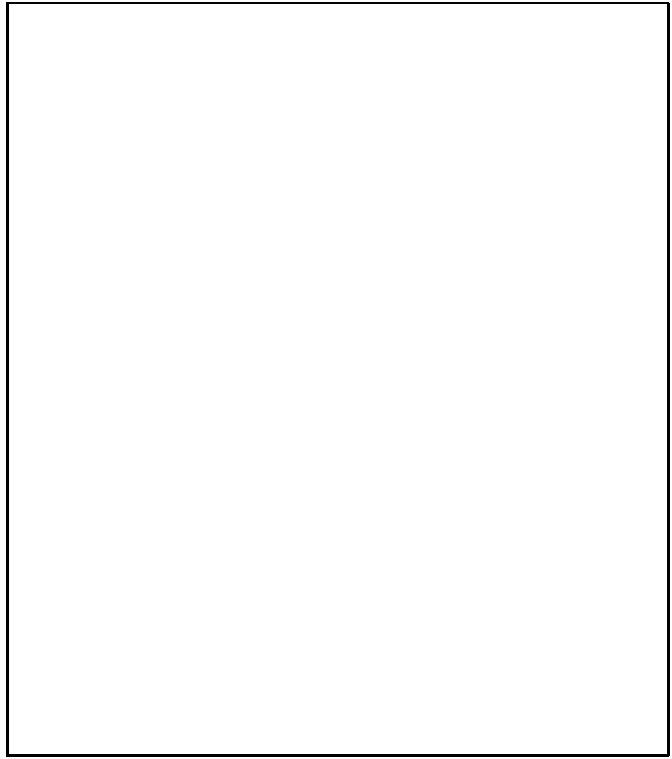

\picplace{10cm}
%\def\epsfsize#1#2{0.7\hsize}
%\centerline{\epsffile{figure3a.ps}}
%\def\epsfsize#1#2{0.7\hsize}
%\centerline{\epsffile{figure3b.ps}}
\caption{Position-velocity diagrams parallel to the minor axis of \n2188:
Cut through {\bf a.} the southernmost filament; {\bf b.} the superbubble
east of the disk and close to the center of \n2188. 
Contour levels are 2,4,...,16 mJy/beam.} 
\end{figure}

Imbedded into the low surface brightness emission surrounding 
the disk of \n2188, HI filaments at higher surface brightness 
can be found emerging from the disk and extending into the halo. 

One very prominent feature 
extending $\sim$1.8 kpc perpendicular to the galaxy's disk is located 
south-west of the disk at $\alpha,\delta(2000) = 06^{h}\,10^{m}\,08^{s}, 
-34^{\circ}\,07'\,20''$.  The position-velocity ($pv$) 
diagram in Figure 3.a shows that the
radial velocity of this ``worm'' increases with distance from the plane. 
The velocity difference compared to the underlying disk reaches up to 
$\sim 25 \kms$. 

Two other spur-like HI structures east of the disk close to the center 
of the galaxy, can be shown to be connected to a superbubble expanding
in the area where the two features leave the disk
($\rm \alpha,\delta(2000) = 06^{h}\,10^{m}\,0\fs5, -34^{\circ}\,06'\,18''$).
Figure 3.b shows a $pv-$diagram along the minor axis at 
$\delta(2000) = -34^{\circ}\,06'\,18''$. 
Thus the velocity structure of the gas between the two HI ``spurs'' 
is displayed. The velocity spread of the HI gas shows clear signatures 
of bubble expansion with an expansion velocity of 
about $v=10 \kms$. The radius of the bubble is $15''$ (575 pc) and its flux is about $0.7$ Jy$\kms$ corresponding 
to $6\times10^6$ M$_{\odot}$.  Following Heiles (1979), an energy
of $1.6 \times 10^{53}$ erg is needed to produce a structure with these
properties.

\subsubsection{The velocity field}

Figure 4 displays a contour plot of the intensity weighted HI velocity
field of \n2188 superposed on its grey scale representation. 
In order to emphasize the large scale structure 
of the velocity field, we present the natural weighted version of the 
data ($25''$ resolution).
In the northern half of the disk of \n2188 the 
velocity field is that of a
highly inclined rotating disk.  In the halo east of the disk
the velocity contours bend towards north. However, in the southern half 
of the disk as well as the halo, the velocity contours are very irregular. 
This is especially true for that part of the galaxy where the most
prominent HI filament extends from the disk into the halo ($\S\ 3.2.1$). 
For $\delta < -34^{\circ}\,08'$ the velocity gradient follows the disk 
bending to the east. Independent of the peculiar gas distribution
(see above), these kinematic disturbances also indicate a large-scale 
perturbation of the neutral gas in \n2188. 

\begin{figure}
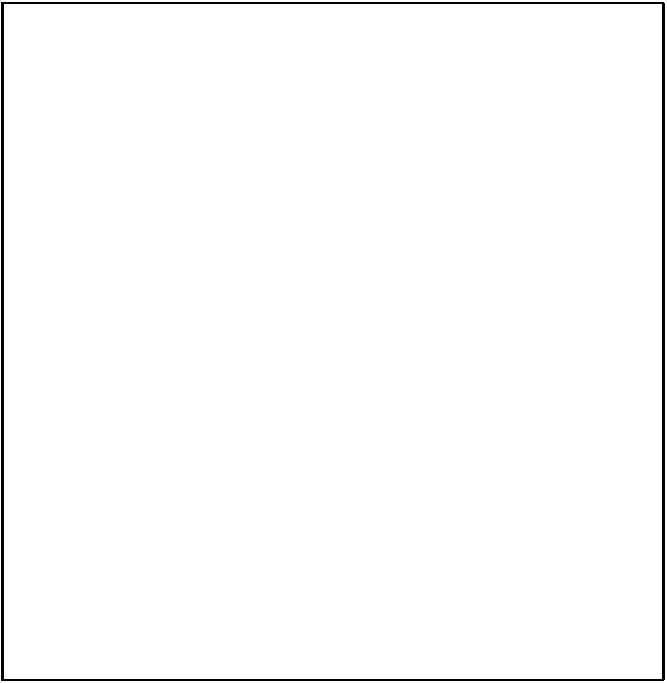

\picplace{9cm}
%\rotate[r]{
%\def\epsfsize#1#2{1.0\hsize}
%\centerline{\epsffile{figure4.ps}}}
\caption{Velocity field of \n2188.  The resolution is $20''$. 
Contour levels are: 700,710,720,...,790 $\kms$.}
\end{figure}

\begin{figure}
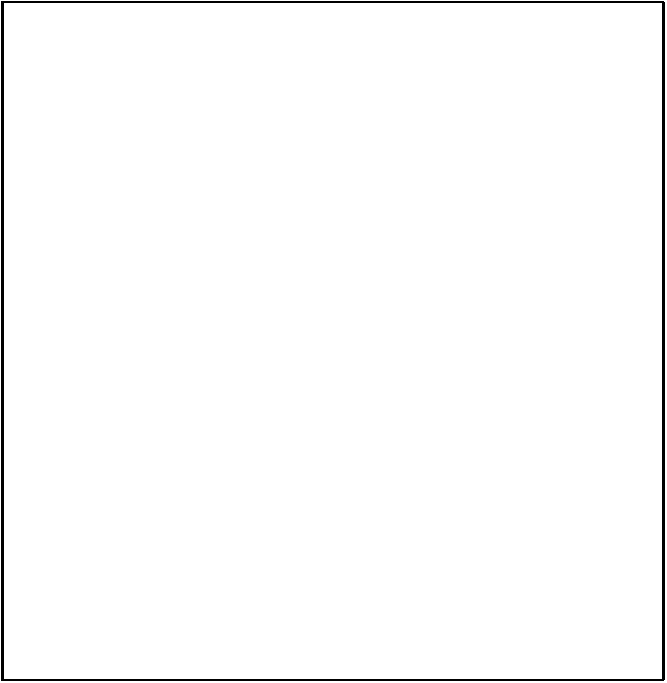

\picplace{9cm}
%\def\epsfsize#1#2{0.9\hsize}
%\centerline{\epsffile{figure5.ps}}
\caption{Position-Velocity diagram along position angle 0$^{\circ}$
at $\alpha= 6^{h} 10^{m} 9\fs9$. Contour levels are: 24,36,48,...,600 mJy/beam.}
\end{figure}

Figure 5 shows a $pv-$diagram along position
angle 0.  This is nearly but not exactly equivalent to 
a $pv-$diagram along the major axis of \n2188 due to the 
bent shape of the galaxy's disk.  The inner $1\farcm5$ of 
the disk are rotating ridgidly.  At $\sim\delta = -34^{\circ}05'30''$
and at $\sim\delta = -34^{\circ}07'00''$ signs for the onset of 
differential rotation can be found.  However, it is quite hard
to describe the $pv-$diagram just by rigid and differential
rotation.  

From the HI spectrum of \n2188 we determine the velocity width at a 20\% level
to be 145$\kms$.  This is consistent with what we find
from Figure 5 and allows us (in combination with the radius for 
rigid rotation) a rough estimate of the total mass in the inner
3.5 kpc radius of \n2188.    
We find $M_{\rm tot}\sim 2.1\times10^9$ M$_{\odot}$.

\subsection{The structure of the ionized gas in \n2188}

In Figure 1 we display our continuum subtracted \ha\/ image of 
\n2188.\footnote{Note that while we will be speaking about ``H$\alpha$'', 
a contribution of about $\sim$20\% to the line emission comes from the 
$\lambda\lambda 6548/6583 {\rm \AA}$ [NII] lines. No correction for
this has been applied.}
The star formation activity which is mostly concentrated in the south
of \n2188, in a region 2 kpc across along 
the major axis, is asymmetric with respect to the galaxy centre.  
Probably the most spectacular features on the \ha\/ image are the 
prominent emission line filaments.  These start in the most conspicuous 
star formation regions in the disk and extend up into the halo of \n2188. Their 
lengths range from $\sim 150$ pc up 
to $\sim 500$ pc, while their typical width is only $\sim40$ pc.
Most of the filamentary emission is found west of the disk. The 
southern HII region complex is also surrounded by an extended envelope 
of diffuse emission line gas with a thickness of $\sim350$ pc.

A comparison of the HI contours and the \ha\/ distribution shows an 
overall correspondence of the HI high surface brightness disk and the 
star forming disk. The disk has a ``banana''-shaped appearance, as
already noted for the HI distribution.

\section{Discussion and conclusions}

\subsection{\n2188: a perturbed galaxy}
The most surprising result of the investigation of the neutral gas
in \n2188 is the large number of irregularities and asymmetries
which are present in the data.
All these peculiarities, i.e., 1) the ``banana'' shape of the disk, 
2) the asymmetric distribution of gas around the main body of \n2188, 
3) the excess optical continuum emission in the northern half
of the disk, and 
4) the disturbed velocity field of \n2188 suggest a large-scale 
(possibly external) disturbance. 

\n2188 is a member of a small group of only three galaxies (Tully 1988). 
Its closest neighbour (ESO\,$364-29$) 
has a systemic velocity only 42 $\kms$ higher than 
that of \n2188, but its projected distance to \n2188 is 180 kpc 
and it is even smaller than \n2188 itself.
If we assume a relative velocity of the two galaxies of 200 $\kms$,
nearly $2\times10^9$ years ($\sim$ 8 rotation periods)
must have passed since their closest passage.  
A relative velocity of 200 $\kms$ is somewhat higher than 129$\pm$22 $\kms$, the
``square root of the average squared rms radial velocity of the galaxies
in a group with respect to its average system velocity'' determined
for 18 loose groups of galaxies by Williams and Rood (1987).
Thus, it seems unlikely that the peculiar gas distribution of \n2188 
can be explained by an interaction with that galaxy. 

The shape of the disk of \n2188 as well as the HI distribution 
are both reminiscent of bow shock phenomena as seen in galaxies
moving into a tenuous intracluster gas (e.g. Cayatte et al. 1994)
and the large scale disturbances of \n2188 can coherently 
be explained by gas stripping processes due to an intergalactic
medium.  In order to investigate if this is a viable process (although
\n2188 is not a member of a galaxy cluster) we will now 
estimate a lower limit for the density of a possible intragroup medium.  
There are basically two processes by which an intergalactic 
medium can remove gas from a galaxy: ram pressure stripping 
and stripping due to transport processes.

In the case of ram pressure by the intergalactic
medium the ram pressure has to be higher than the local surface gravity
of the interstellar gas:  
$\rho_{\rm IGM} v^2 > 2\pi G \sigma_{\rm tot} \sigma_{\rm gas}$
(Gunn \& Gott 1972).  $\rho_{\rm IGM}$ is the density of the
intergalactic medium. 
$\sigma_{\rm tot}$ and $\sigma_{\rm gas}$ are the total surface densities 
and the gas surface density respectively. They can be estimated from the
HI data.  $\stot$ we calculate from the maximum velocity and 
the corresponding radius to be $2.2 \times 10^8$ M$_{\odot}$ kpc$^{-2}$.  
$\sgas$  we estimate 
assuming that the HI mass in the disk of \n2188 is distributed evenly over
the HI disk area which has 
a radius of $\sim 4.3$ kpc.  We take only
the disk gas (2$\times$10$^8$ M$_{\odot}$) under consideration
and get $\sgas = 3.4 \times 10^6$ M$_{\odot}$ kpc$^{-2}$.
Finally, we again assume v$\sim 200$ $\kms$ for 
the velocity of \n2188 perpendicular to its plane.
All these assumptions yield $\rigm > 5 \times 10^{-3}$ cm$^{-3}$.  

Transport processes can cause stripping
of gas with a rate that in some cases exceeds that caused by
ram pressure stripping (Nulsen 1982).  
Nulsen found that the mass loss rate 
of a galaxy due to thermal conduction and viscous stripping
can be expressed as  {\it \.{M}}  $= \pi r^2 v \rho_{\rm IGM}$.  
We assume that the amount of mass lost by \n2188 
is comparable to the fraction of disk gas that resides 
in an area comparable to the northern gas deficient part 
of \n2188. This area is roughly equivalent to a quarter of a ring 
with a thickness of 1.7 kpc ($0\farcm76$). 
Using $\sgas$ from above, we obtain $M > 4.7 \times10^7$ M$_{\odot}$.
For mass loss timescales longer than half a rotation period
of \n2188 a gas deficiency should be observed over the 
entire galaxy.  Therefore, a reasonable timescale
for the mass loss probably is half a rotation period of \n2188
($\sim 1.2 \times 10^8$ years).  Using $r=4.3$ kpc and $v=200$ $\kms$
we find $\rigm \sim 1\times10^{-3}$ cm$^{-3}$.

Consequently, if the appearance of \n2188 is due to
gas stripping processes $\rigm$ has to be at least $1\times10^{-3}$
cm$^{-3}$.  This is comparable to the densities of the intergalactic
medium in galaxy clusters (Sarazin 1992).  Since \n2188
is not a cluster member it seems unlikely
to us that the morphology of \n2188 is due to gas stripping
processes.  

However, it is possible that \n2188 interacted or is interacting
with an intergalactic HI cloud. In this scenario, both the peculiar
gas dynamics and the absence of a visible (stellar) interaction
partner could be explained.  In fact, the morphology of
the neutral gas in \n2188 strongly reminds of similar features
in IC\,10 where they have been interpreted as 
an interaction with an HI plume (Shostak \& Skillman 1989).

\subsection{The disk-halo connection of \n2188}

Having HI as well as \ha\/ data available also allows us to study
the interstellar disk-halo connection of \n2188.   \n2188 is certainly 
remarkable for its prominent features of diffuse ionized gas (DIG).  
Figure 1 shows
that there is a clear correlation between star formation processes
in the disk and the occurence of DIG in the halo of \n2188.  
The HI data show also that peculiar filaments of neutral gas 
extending from the disk into the halo of \n2188 exist.  
Similar to the HI features in NGC\,3079 (Irwin 1990)
these filaments can be interpreted in terms of Heiles shells.
Then their presence suggests that superbubble expansion
is an important process for the disk-halo interaction of \n2188.

Comparing the HI and the H$\alpha$ data we do not find 
a correlation between DIG filaments and superbubble features.  In particular
the giant superbubble east of the center of \n2188 does not show 
prominent DIG emission.  
This is unlikely to be due to dust absorption, since \n2188 
is a late type galaxy and since the HI features are found above the 
disk where the column densities are low. 
The absence of such a correlation can very well be explained by an
age sequence: during the lifetime of an OB superassociation 
diffuse ionized structures are produced in its vicinity
and expand into the halo of the galaxy where
densities are lower.  As the cluster ages
the ionized structures recombine quickly.  However, HI structures
which have been formed due to the combined energy input of stellar
winds and supernovae can still be observed 
even when all the diffuse \ha\/ emission has vanished.

\acknowledgements{We are thankful to the NRAO for providing their
facilities during the process of data reduction.  We also thank
E. Brinks, E.D. Skillman and K.S. de Boer for useful discussions.  H.D. was
supported by the DFG Graduiertenkolleg Magellansche Wolken and additional
travel support was granted under De 385/8-1.
This research has made use of the NASA/IPAC Extragalactic Database 
(NED), which is operated by the Jet Propulsion Laboratory, Caltech, 
under contract with the National Aeronautics \& Space Administration. 
}

\end{document}